\documentclass[a4paper]{jpconf}
\usepackage{graphicx}
\usepackage{wrapfig}

\begin{document}
\title{TAU-4 installation intended for long-term monitoring of a half-life value of the $^{212}$Po\footnote{The 4th International Conference on Particle Physics and Astrophysics (ICPPA-2018)}
}

\author{E.N.Alexeev$^{1}$, Yu.M.Gavrilyuk$^{1}$, A.M.Gangapshev$^{1,3}$, A.M.Gezhaev$^{1}$, V.V.Kazalov$^{1}$, V.V.Kuzminov$^{1,3}$, S.I.Panasenko$^{2}$, O.D. Petrenko$^{2}$, S.S.Ratkevich$^{2}$}

\address{$^1$ Baksan Neutrino Observatory INR RAS, Russia}
\address{$^2$ V.N.Karazin Kharkiv National University,Ukraine} \address{$^3$ H.M.Berbeckov Kabardino-Balkarian State University,Russia}

\ead{bno\_vvk@mail.ru} 

\begin{abstract}
Description of the TAU-4 installation intended for long-term monitoring of the half-life value $T_{1/2}$ of the $^{212}$Po is presented. Natural thorium used as a source of the mother’s chain. The methods of measurement and  data  processing are described. The comparative results of short test measurements carried out in the ground (680 h) and underground (564 h) laboratories are given. Averaged value $T_{1/2}$ =$294.09\pm0.07$ ns of the $^{212}$Po half-life has been found  for the ground level data set similar one for the underground data set.  The solar-daily variations with amplitudes $A_{\rm So}=(11.7\pm5.2)\times10^{-4}$ for the ground data and  $A_{\rm So}=(7.5\pm4.1)\times10^{-4}$ for the underground one  were found in a series of $\tau$ values.
\end{abstract}

\section{Introduction}
Experimental research of the half-life time stabilities of $^{214}$Po (October 2012 - May 2015, TAU-1 and TAU-2 setups) \cite{r1,r2,r3} and $^{213}$Po (Julay 2015 up to now, TAU-3 setup) \cite{r4} carry out at the Baksan Neutrino Observatory of the INR RAS. A half-life defines as a result of analysis of the decay curves constructed from a set of life-time values of separate nuclei of the exploring isotope. Delays between the birth of the nuclear ($\beta$-particle from the $^{214}$Bi($^{213}$Bi) decay + $\gamma$-quantum) and  it’s decay ($\alpha$-particle from the $^{214}$Po($^{213}$Po) decay) measured to define this parameter.

The objects of further analysis are time series of $\tau$ values with different time steps.
Annular variation with an amplitude $A=(9.8\pm0.6)\times10^{-4}$, solar-daily variation with an amplitude $A_{\rm Sol}=(5.3\pm0.3)\times10^{-4}$, lunar-daily variation with an amplitude $A_{\rm L}=(6.9\pm2.0)\times10^{-4}$ and sidereal-daily variation with an amplitude $A_{\rm Sid}=(7.2\pm1.2)\times10^{-4}$ have been detected in the $^{214}$Po half-life series ($T_{1/2}=163.47\pm0.03$ $\mu$s).
Annular variation with an amplitude $A=(3.2\pm0.4)\times10^{-4}$, solar-daily variation with an amplitude $A_{\rm Sol}=(5.3\pm1.1)\times10^{-4}$, lunar-daily variation with an amplitude $A_{\rm L}=(4.8\pm2.1)\times10^{-4}$ and sidereal-daily variation with an amplitude $A_{\rm Sid}=(4.2\pm1.7)\times10^{-4}$ have been detected in the $^{213}$Po ($T_{1/2}=3.705\pm0.001$ $\mu$s) half-life series.

Assumptions about the connection of the obtained variations with changes in the operating characteristics of the detectors or recording equipment caused by the influence of variable environmental factors were checked. So far, none of the experiments have confirmed the existence of such an effect.
However, such a possibility cannot be excluded completely because of a large amount of the environmental active factors (air pressure, humidity, and temperature; air ionization level; geomagnetic field and so on) and a complication of its detection at a needed level of sensitivity for a relatively low count rate of the useful events.

Useful event count rate (correlated delayed coincident pulses) on the separate installation depends on the source activity, the dead time of the used digital oscilloscope and the allowed level of an accidental coincidence background.
The relative magnitude of the background of random coincidences is related to the full activity of all the daughter members of a number of the parent source and the duration of the time interval for the selection of delayed coincidences.
A value of the last parameter increases quadratically with a source activity rise and linearly with a selection time interval rise. A useful event count rate increases linearly with a source activity rise. As far as, a half-life value is determined as a result of computer separation of a delays curve on a decay exponent and a flat background substrate, a relative error  in determining the half-life value is related to the value of the background level and increases with its growth and could overlap possible variations.

Maximum permissible level of accidental coincidences reached a relatively low activity of the $^{226}$Ra source for the case of the $^{214}$Po which has a relatively large half-life value. For example, a background is $\sim 1$\% for the 12 s$^{-1}$ useful events count rate. Limitation connected with an accidental coincidences background is practically unessential for the $^{213}$Po ($^{229}$Th mother source). A problem of a necessity of the source purification from the $^{226}$Ra appears in this case because of radium chain can create correlated delayed coincidence background. A computer separation of a delays curve on a decay exponent and a flat background base will become with errors because of a real background will contain nondescript part of the exponential background.

Digital storage oscilloscopes (DO) used for the pulses recording have a significant dead time when applying for a recording of double-pulse events with an ``on-line'' rejection of single-pulse events. An ultimate recording rate is $\sim 50$ s$^{-1}$ in such mode. Pulses from the massive NaJ scintillation detector used for a start of the DO record for the purpose to eliminate essential part of the single pulses. The NaJ detector register $\gamma$-quanta accompanied the decays of the mother Bi-isotopes. The NaJ detector is placed in the underground laboratory and surrounded by a low background shield to decrease the number of false starts. Due to this, essential increase of the useful events counts rates ($>100$ s$^{-1}$) could be reached by increasing a number of simultaneously operated underground setups. A task is highly laborious and expensive. It seems that the main difficulties on the path to the high rate setup could be excluded by using the $\alpha$-active isotope $^{212}$Po with the half-life $(294.7\pm1.0$) ns \cite{r5} as an object of observation.

\section{Setup description}

Isotope $^{212}$Po is a member of the $^{232}$Th radioactive series \cite{r6}: $^{232}$Th ($T_{1/2}=1.4\times10^{10}$ y, $\alpha$)$\rightarrow$ $^{228}$Ra ($T_{1/2}=6.7$ y, $\beta$) $\rightarrow$ $^{228}$Ac ($T_{1/2}=6.1$ h, $\beta$) $\rightarrow$ $^{228}$Th ($T_{1/2}=1.9$ y, $\alpha$) $\rightarrow$ $^{224}$Ra ($T_{1/2}=3.6$ d, $\alpha$) $\rightarrow$ $^{220}$Rn ($T_{1/2}=55.3$ s, $\alpha$) $\rightarrow$ $^{216}$Ро ($T_{1/2}=0.158$ s, $\alpha$) $\rightarrow$ $^{212}$Pb ($T_{1/2}=10.6$ h, $\beta$) $\rightarrow$ $^{212}$Bi ($T_{1/2}=60.5$ min, $\beta$ (64\%) + $\alpha$ (36\%)) $\rightarrow$ $^{212}$Po ($T_{1/2}=2.9\times10^{-7}$ s, $\alpha$) + $^{208}$Tl ($T_{1/2}=3.05$ min, $\beta$) $\rightarrow$ $^{208}$Pb (stable).
It is seen from the presented data that long-lived isotopes which could be the quasi-stable generator of $^{212}$Po nuclei except
$^{232}$Th are absent in the series. A total equilibrium activity of the source will determine by a sum of all series member activities. One correlated delayed coincidence from the $^{212}$Bi-$^{212}$Po pair decay falls on 13.7 ordinary decays in this case.

An activity of the $^{212}$Po at a level of 20 Bq at the series equilibrium condition contained in the $1.6\times10^{-2}$ g of the Th(NO$_3$)$_4$ salt with a volume of $1.7\times10^{-3}$ cm$^3$ (9.2 g$\times$cm$^{-3}$ density). The source area will be 17.3 cm$^2$ if it is assumed to be 1 micron thick. The source was prepared by an evaporation of a solution with a required salt amount on a surface of a mylar film with the 2.5 $\mu$m thickness and 4 cm diameter. The source was covered from above by a similar film. The source was placed between two discs of a plastic scintillator (PS) with the 2 mm thickness and 40 mm diameter and glued by an epoxy resin along the circle. The PS with a source is placed on an edge at the bottom of a cylinder with a diameter of 50 mm and a depth of 100 mm with walls covered with a reflective film VM-2000. The open end of the  cylinder is viewed by the PMT Hamamatsu R12699 SEL photomultiplier.
The assembly has been placed inside a stainless still cylindrical case with 98 mm inner diameter, 200 mm length and 2 mm wall thickness. The recording setup consists of a digital oscilloscope (LA-n1USB) with a digitization frequency of 500 MHz (2 ns per bin) connected to a personal computer (PC).

\begin{wrapfigure}[13]{l}{0.5\textwidth} \vspace{0.5pc}
\includegraphics[width=19.0pc,angle=0]{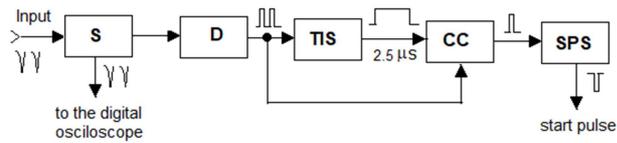}
\caption{\label{f1} {\small Block-diagram of the DO start pulse former intended for selection of the delayd pulses.
(S)  - splitter, (D) - discriminator, (TIS) - time interval shaper, (CC) - coincidence circuit,
(SPS)  -  starting  pulse shaper.}
}
\end{wrapfigure}
A method of eliminating single pulses from the full flow of pulses to the input of the DO has been proposed for an order to increase the efficiency of the registration of double correlated pulses.
For this purpose, a scheme for generating a startup DO pulse was created only in response to the appearance of two delayed pulses in a given time gap.
The block diagram of the shaper is shown in Fig.\ref{f1}.
Trigger pulse shaper consists of a signal splitter (S), a fast discriminator (D), time interval shaper (TIS),
coincidence circuit (CC) and starting pulse shaper (SPS).
Pulses from the PMT anode come to the splitter input and divide into two. Pulses from the S-outputs come to the DO-input and to the D-input.
The discriminator forms short logic pulses if the anode pulse
amplitude exceeds 30 mV. The TIS triggers by the first pulse and
forms pulse with $\sim 2.6$ $\mu$s duration. The CC opens for this time and passes the second pulse if it comes in this time gap.
The CC does not pass the first pulse because of the TIS triggered by its falling edge. The SPS generates a 50 ns duration output pulse for a triggering of the DO record if the second input pulse appears at this $\sim 2.6$ $\mu$s.

The DO records information contained in the ``prehistory'' which is the (-2.6 $\mu$s)-time range before the triggering pulse appearing ($\sim 8$ half-lifes of the $^{212}$Po). It was found during a preliminary measurement that the used PMT has no afterpulses which can distort a form of delays distribution at a short time region, therefore a long time series of measurements was started at this installation, which was called TAU-4.

\section{Results of preliminary measurements}
\begin{figure}
\begin{center}
\includegraphics[width=11pc,angle=270]{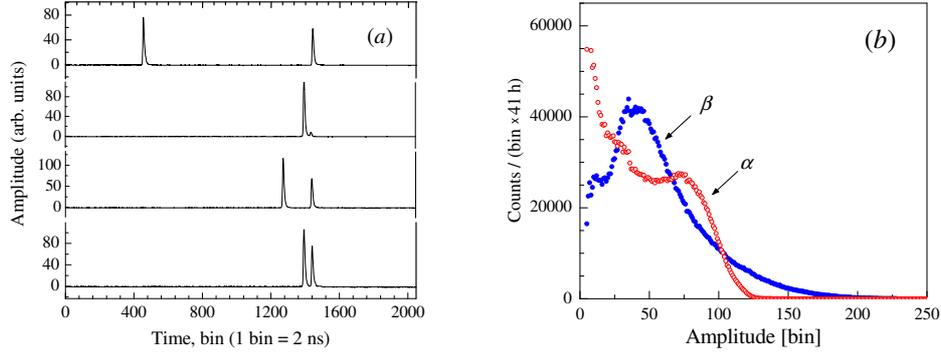}
\caption{\label{f2_3} {\small
{\it a}: examples of selected work events with two correlated pulses; {\it b}: Amplitude spectra of $\beta$- (blue) and $\alpha$-particles (red).}}
\vspace{-1.0pc}
\end{center}
\end{figure}

The recording rate of events with double delayed pulses to the memory of the PC that controls the operation of the DO was $\sim 20$ s$^{-1}$.
An  examination of the recorded information shows that practically all events contain two pulses. Four  samples of the working recorded frames are shown on Fig.\ref{f2_3}$a$. The $\alpha$-particle pulses which generate trigger pulses started the DO recording cycles are positioned on the $\sim 1470$ time bin. First pulses from the $\beta$-particle which initiate the former working  algorithm is in the ``prehistory''. The position of this pulses varies from frame to frame.  An array of the data about first and second pulses amplitudes and appearing time is formed in process of ``off-line'' processing of the recorded frames.
A spectrum of the $\beta$-particles of the $^{212}$Bi decay formed from the amplitudes of the first pulses is shown on Fig.\ref{f2_3}$b$.
A spectrum of the $\alpha$-particles of the $^{212}$Po decays is also shown there. A data collection time for this plotting was 41 h.

\begin{figure}
\begin{center}
\includegraphics[width=10pc,angle=270]{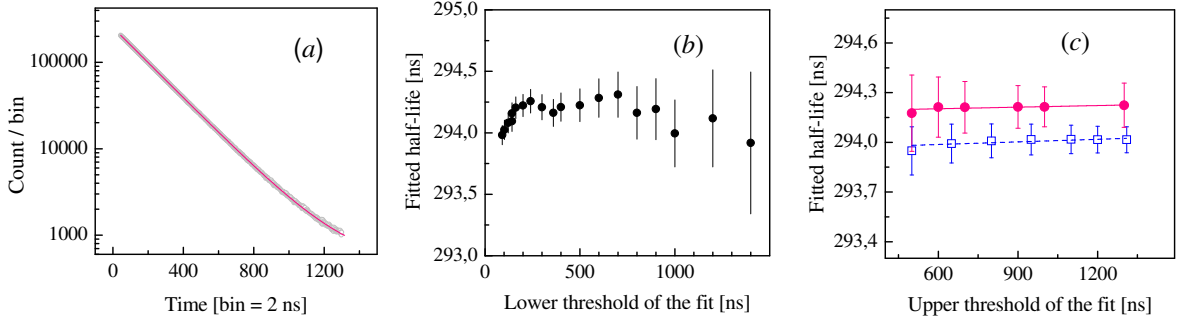}
\caption{\label{f4_5} {\small
$a$: the decay curve of $^{212}$Po accumulated over 680 h;
$b$: the dependence of $\tau$ values on the magnitude of the lower threshold of delays at fixed upper threshold (2600 ns); $c$: dependence of $\tau$ on the magnitude of the upper threshold for two values of the lower threshold - 100 ns and 300 ns  - red and blue dots respectively.
}}
\vspace{-1.5pc}
\end{center}
\end{figure}

The initial measurement was started 30 May 2018 in a laboratory on the second floor of a ground building. The setup was surrounded 10 cm lead shield to decrease a radioactive background. A delay between the first and the second pulses corresponds to a life-time of a separate $^{212}$Po nucleus. A set of delay values at a stated period are used for the construction of a decay curve. A decay curve collected at 680 h is shown of Fig.\ref{f4_5}$a$.
None selection on amplitudes of the $\beta$- or $\alpha$-spectra was not used at its construction. The curve was approximated in the 150-2600 ns range of delays (75-1300 bins of time) by a function
\[
f(t)= A \cdot exp(-ln(2)∙t/\tau)+b.
\]
A half-life value $\tau\equiv T_{1/2}=294.09\pm0.07$ ns was found. This value in the limits of errors coincides with values $[294.7\pm1.0$ ns$]$ and $[293.9\pm 1.0(stat.) \pm 0.6(syst.)$ ns$]$ obtained in the work \cite{r5} and \cite{r7} but slightly less than the value $[298.8 \pm  0.8(stat.) \pm 1.4(syst.)$ ns$]$ obtained in the \cite{r8}.

A measurement was continued after that in the low background laboratory DULB-4900 of the Baksan Neutrino Observatory of the INR RAS \cite{r9}. A half-life value $\tau = 294.07\pm0.08$ ns has been  found for the decay curve collected at 564 h. One can see that the half-life values obtained in the ground and underground conditions are coincided. A conclusion could be made that values of various environmental parameters which essentially different in values and stability in the both setup locations do not affect on the half-life value which averaged at the $\sim$one month period. A half-life value for a specific setup could depend on a delay low threshold value of an analyzed decay curve as it follows from the \cite{r7}. We have checked this observation on our data.
A dependence of a half-life value on a low delay threshold value for the fixed 2600 ns upper threshold is shown on Fig.\ref{f4_5}$b$
for a sum data set.  Dependencies of a half-life value on an upper delay threshold for the two values of a low threshold (100 ns and 300 ns) are shown on Fig.\ref{f4_5}$c$.
A $\tau$-value does not depend practically on an upper delay threshold but depends on low delay threshold in a region of 90-160 ns and varies at 0.07\%.
A small distortion at low delays could be explained by an affection of features of the trigger signals generation at low delays and possible amplitude dependence of the former sensitivity on a pulse amplitude. Due to this, some pulse combination
passed the former at higher delays could not pass selection condition at low delays.

\begin{figure}
\begin{minipage}{12.5pc} \vspace{-0.40pc}
\includegraphics[width=12.3pc]{{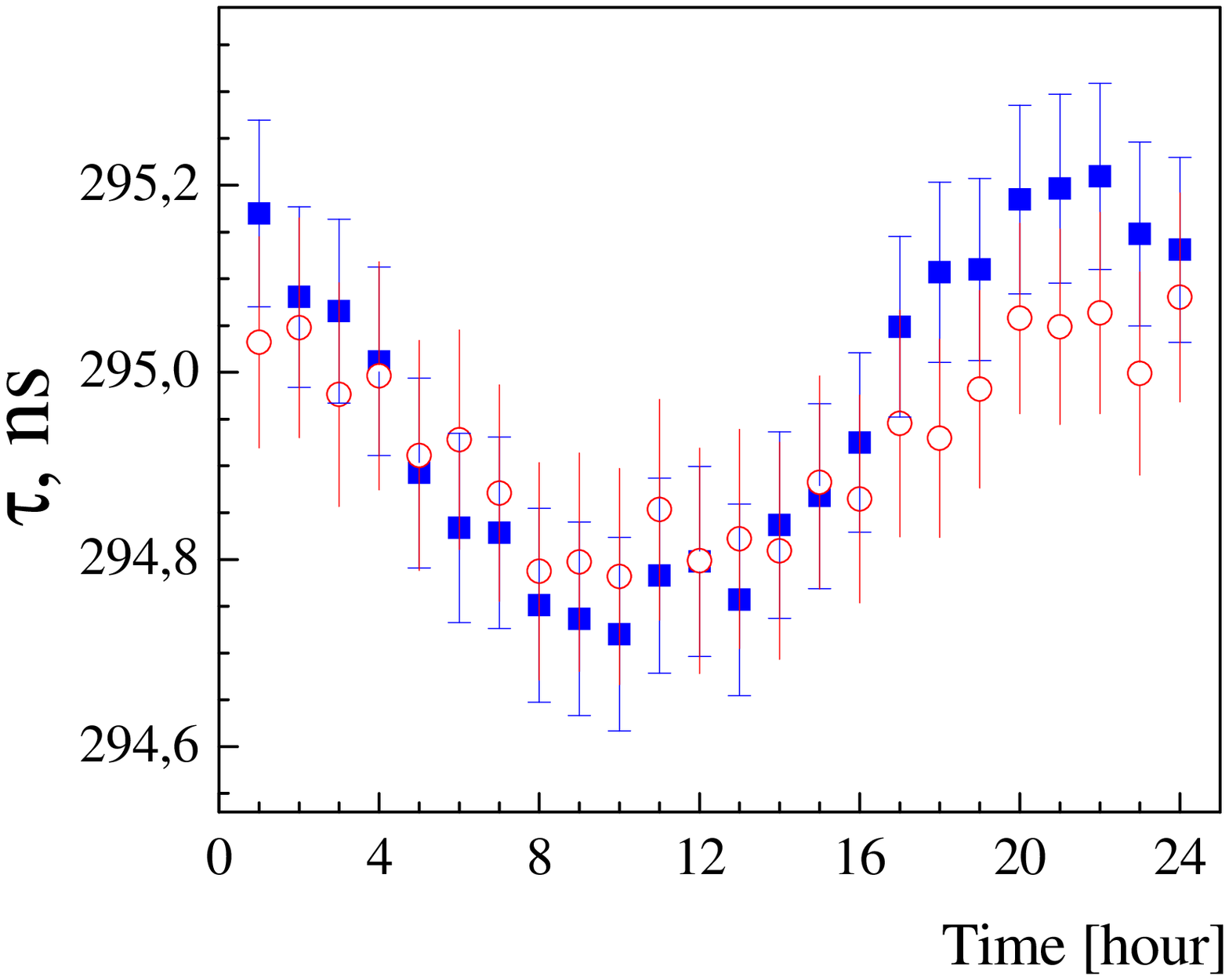}} \vspace{-0.5pc}
\caption{\label{f6}{\small Dependences of the half-life on the solar day time for two data sets: blue dots - ground-based measurements; red dots - underground measurements.}}
\end{minipage} \hspace{0.6pc}%
\begin{minipage}{24pc} \hspace{-0.3pc}
\vspace{0.70pc}
\includegraphics[width=10pc,angle=270]{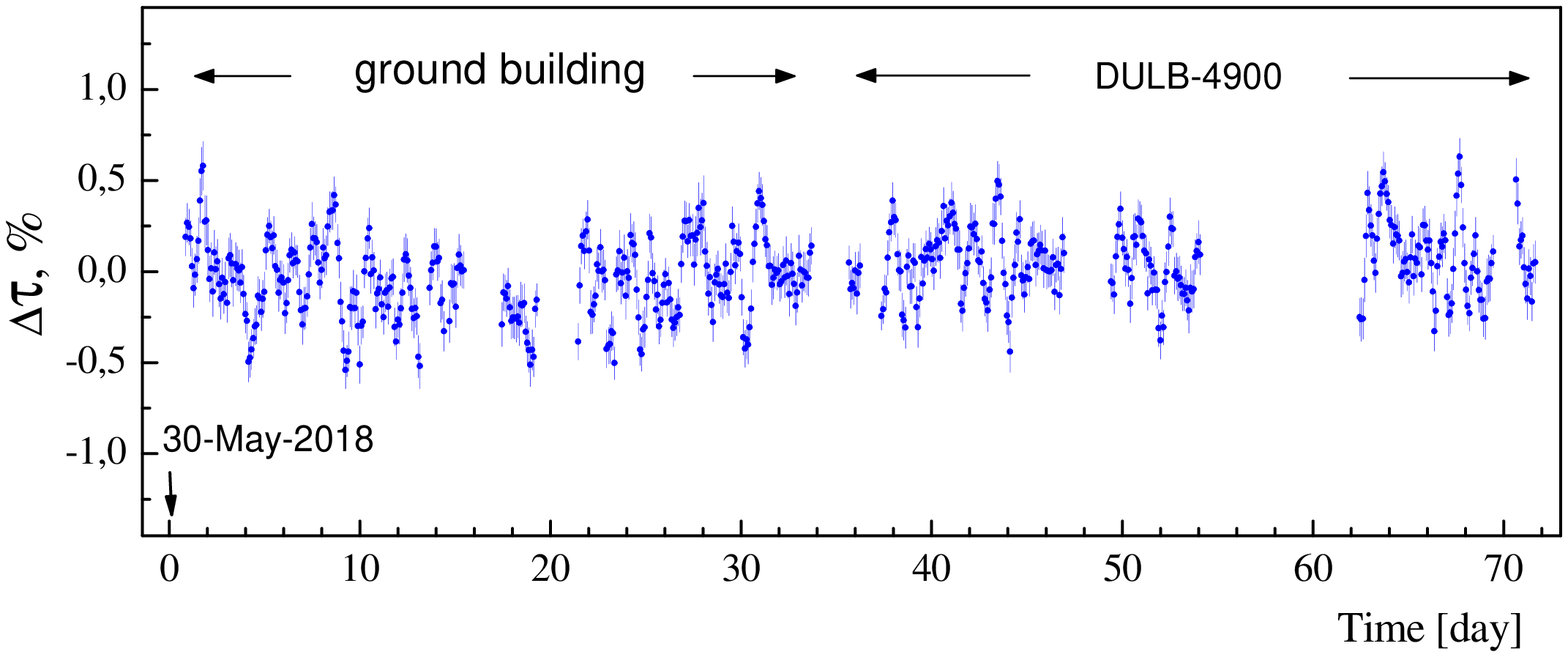} \vspace{-1.5pc}
\begin{center}
\caption{\label{f7}{\small
Time dependence of the deviations of the half-life of $^{212}$Po
relative to its average value over the total measurement period in the ground (from 30-May to 03-Jul 2018 y)  and underground (from 05-Jul to 10-Aug 2018 y) laboratories. Experimental points with errors were obtained from the decay curves accumulated over 12 h in increments of one hour.}}
\end{center}
\vspace{-2.0pc}
\end{minipage}
\end{figure}

Distributions of half-life values on a solar daytime has been  constructed to search for possible day variations. The dependence of $\tau$-value for the ground data is shown on Fig.\ref{f6} by squares and the one for the underground data is shown on Fig.\ref{f6} by circles.
The inner moving-average method (IMA) was used to improve the  statistical precision of a $\tau$-value determination.
Data from the time interval 0-12 h in all measured days are collected to construct a decay curve. The obtained half-life value enters to the first point of the graph. A time interval shifted for a one hour (1-13 h) to determine the second half-life value and so on. Adjacent data points obtained by this method differ for 1/12 only. An IMA procedure integrates a real function over half of the period, practically. It is seen from Fig.\ref{f6} that $\tau$-dependencies could be approximated by sine functions. An amplitude of a real function should be reconstructed by multiplication of an approximation sine amplitude on $\pi/2$. A phase (the point where the amplitude equal to zero and will grow with the argument increase) will move at 6 h
$[24{\rm h}\cdot (2\pi)/(\pi/2)]$ from the 17 h point to the 23 h point.

Error bars shown on the Fig.\ref{f6} have similar values and present the systematic errors, practically. This error appears if one uses a sum function $ f=A \cdot exp(-x)+b$ as an approximation curve and multiply reduce if a pure exponent uses as an approximation.

A value of a reconstructed amplitude of the solar-day variation is equal to $A_{\rm So}= (7.5 \pm 1.7(stat.)\pm3.3(syst.)\times10^{-4} =(7.5\pm4.1)\times10^{-4}$ for the ground data and is equal to $A_{\rm So}=(11.7\pm5.2)\times10^{-4}$ for the underground data. A using of a pure exponent for an approximation of a decay curve allows one to get a presentation about a behavior in time of the initial $\tau$-values because of error values became much smaller than amplitude variations. At the same time, the absolute $\tau$-values insignificantly differ from the values obtained if a sum function is used for the approximation. The amplitudes and shapes of the current variation will be almost the same. Behavior in time of a normalized half-life value of the $^{212}$Po is shown on Fig.\ref{f7} for the total measurement time interval. Each hour point has been obtained from a decay curve collected at $\pm6$ hours distance centered on around the chosen hour. Time intervals of ground (30 May - 03 July 2018 y) and underground (05 July - 10 August 2018 y) measurements are marked by arrows on the Fig.\ref{f7}.

More accurate conclusions about parameters of the variation will be hasty because of a short measuring time and lacunas in the data series connected with the setting of setup.

\section{Discussion of results}
Values of the $^{212}$Po experienced statistically significant variation in time. The amplitude of daily variations can reach 1\%
and more for the moment. It is could be assumed that such variations could be present in the data measured with the $^{212}$Po at the others setups because of up to now it does not find an unambiguous connection of the variations measured with the TAU-4 setup and variations of any environmental parameters. A comparison of the results for the $^{214}$Po, $^{213}$Po, and $^{212}$Po shows that averaged amplitudes of its solar-daily variations are in the region of $\sim 7\times10^{-4}$. Its half-lives relate as
(556):(12.6):(1.0). A set of possible channels of generation of half-life variations caused by the climate factors could be excluded taking these facts into account.
The absence of the amplitude dependence of the variation on the half-life value suggests that the nature of the changes is not associated with possible temporary changes
characteristics of the detector or electronics.
Possible effects should be identical in the absolute values for all listed isotopes and a relative value of variation should increase with a decreasing of a half-life. A search for a possible source of variations should be focused on process having a similar form for all isotopes and not depends on a half-life.
An instability of an amplification coefficient including one of a PMT and electronics or DO circuits could be such source. Variations of the signal amplitudes under the action of microclimatic  factors could  change a real averaged low amplitude threshold in the $\beta$- and $\alpha$-spectra of the events used to construct a decay curve. It could cause variation of a half-life value if there are exists any dependence of a half-life value on an amplitude of $\beta$- and $\alpha$-pulses. An investigation has started in this obtained direction.

\section{Conclusions}
Description of the TAU-4 installation intended for long-term monitoring of the half-life value $T_{1/2}$ of the $^{212}$Po is presented. Natural thorium is used as a source of the mother's chain. The methods of measurement and processing of collected data are reported. Short testing measurements have been made in the ground building (680 h) and underground laboratory (564 h). Averaged value $T_{1/2}=294.09\pm0.07$ ns of the $^{212}$Po half-life has been found for the ground level data set similar one for the underground data set. The solar-daily variations with amplitudes $A_{\rm So}=(11.7\pm5.2)\times10^{-4}$ for the ground data and  $A_{\rm So}=(7.5\pm4.1)\times10^{-4}$ for the underground one  were found in a series of $\tau$ values. Comparison of the $^{214}$Po, $^{213}$Po, and $^{212}$Po half-life values solar-daily amplitude variations shows that averaged amplitude variations are in a region of $\sim 7\times10^{-4}$. This fact allows one to make a conclusion that hypothetic setup time characteristic variations under the influence of environmental variable factors could not be a reason of the half-life amplitude solar-daily variations.
Investigations are continuing. \\

{\bf {Acknowledgments}} \\
The authors express deep gratitude to the V.I.~Volchenko and A.F.~Janin for the development and making of the start pulse former electronic modules.
This work was performed in accordance with a research plan of the Institute for Nuclear Research of the Russian Academy of Sciences with support by the grant of the Program of the Presidium of the Russian Academy of Sciences ``Physics of  Basic  Interactions and Atomic Technologies''.

\end{document}